\documentclass[prl
,superscriptaddress,showpacs,nofootinbib,%
tightenlines]{revtex4}
\usepackage{epsfig}
\usepackage{amssymb}
\newcommand{\beq}{\begin{equation}}
\newcommand{\eeq}{\end{equation}}
\newcommand{\beqa}{\begin{eqnarray}}
\newcommand{\eeqa}{\end{eqnarray}}

\newcommand{\ben}{\begin{displaymath}}
\newcommand{\een}{\end{displaymath}}
\newcommand{\be}{\begin{equation}}
\newcommand{\ee}{\end{equation}}
\newcommand{\bea}{\begin{eqnarray}}
\newcommand{\eea}{\end{eqnarray}}
\begin{document}
\title{Renormalization of the three-boson system with short-range interactions revisited}
\author{E.~Epelbaum}
 \affiliation{Institut f\"ur Theoretische Physik II, Ruhr-Universit\"at Bochum,  D-44780 Bochum,
 Germany}
\author{J.~Gegelia}
\affiliation{Institute for Advanced Simulation, Institut f\"ur Kernphysik
   and J\"ulich Center for Hadron Physics, Forschungszentrum J\"ulich, D-52425 J\"ulich,
Germany}
\affiliation{Tbilisi State  University,  0186 Tbilisi,
 Georgia}
\author{Ulf-G.~Mei\ss ner}
\affiliation{Helmholtz Institut f\"ur Strahlen- und Kernphysik and Bethe
   Center for Theoretical Physics, Universit\"at Bonn, D-53115 Bonn, Germany}
 \affiliation{Institute for Advanced Simulation, Institut f\"ur Kernphysik
   and J\"ulich Center for Hadron Physics, Forschungszentrum J\"ulich, D-52425 J\"ulich,
Germany}
\author{De-Liang Yao}
 \affiliation{Institute for Advanced Simulation, Institut f\"ur Kernphysik
   and J\"ulich Center for Hadron Physics, Forschungszentrum J\"ulich, D-52425 J\"ulich,
Germany}

\begin{abstract}

We consider renormalization of the three-body scattering problem in low-energy effective field theory of self-interacting scalar particles by applying time-ordered perturbation theory to the manifestly Lorentz-invariant formulation. The obtained leading-order equation is perturbatively renormalizable and non-perturbatively finite and does not require a three-body counter term in contrast to its non-relativistic approximation.

\end{abstract}

\maketitle

At very low energies in effective field theory (EFT) for nuclear systems one can
integrate out all particles except the nucleon. At
leading order (LO) of the nucleon-nucleon interaction one is left with a constant contact interaction effective potential. 
A rather non-trivial problem of renormalization arises when one tries
to use this contact interaction in a standard non-relativistic calculation of the doublet channel neutron-deuteron scattering \cite{Bedaque:1998kg,Bedaque:1998km,Bedaque:1999ve}. 
Exactly the same problem occurs in the low-energy non-relativistic EFT of self-interacting scalar particles. 
	In particular, to obtain the LO amplitude for a boson scattering off a two-boson
bound state, an infinite number of diagrams need to be summed up.  
When one tries to achieve this by using an integral equation, one finds that it does not have an unique solution. 
Because of this the solution to the regularized equation does not converge to 
a fixed limiting value as the cutoff is removed.  Bedaque {\em et al.}  argued that the introduction
of a three-body force is a necessary and sufficient condition
to eliminate this cut-off dependence \cite{Bedaque:1998kg,Bedaque:1998km,Bedaque:1999ve}. While there has been an attempt to resolve this problem without introducing a three-body force \cite{Blankleider:2000vi}, 
the conclusions of Refs.~\cite{Bedaque:1998kg,Bedaque:1998km} are commonly accepted, for a review see e.g. Ref.~\cite{Braaten:2004rn}.

Lorentz invariance is a fundamental symmetry underlying any EFT of
particle and nuclear physics. At low energies it is useful and
convenient to expand physical quantities in inverse powers of the
speed of the light. This non-relativistic expansion of physical
quantities can be also reproduced by first expanding the effective
Lagrangian and subsequently calculating the physical
quantities. However, on top of those terms which are obtained from the non-relativistic expansion of the effective Lagrangian, 
one needs to add extra terms which take care of the non-commutativity of the non-relativistic expansion and loop integration (i.e. quantum corrections), see e.g. Refs.~\cite{Bernard:1992qa,Gegelia:1999gf}. 
These additional terms have structures already 
present in the effective Lagrangian and, therefore, they only lead to
the shifts of the coupling constants. Provided that the low-energy
couplings are properly matched, the non-relativistic approach exactly
reproduces the expansion of the Lorentz invariant results wherever
this expansion makes sense. Note, however, that the ultraviolet
behaviour of loop diagrams is qualitatively different in
Lorentz-invariant and non-relativistic theories and, therefore, the
shifts of low-energy coupling constants, which guarantee the
equivalence between the two approaches, usually lead to qualitatively
different behaviour of bare couplings as functions of the cutoff
parameter in non-relativistic and the Lorentz invariant effective
Lagrangians. For example, one-pion exchange (OPE) potential is
perturbatively renormalizable in Lorentz-invariant formulation while
non-renormalizable in non-relativistic theory
\cite{Epelbaum:2012ua}. While one can always add a sufficient number
of counter term contributions to any finite number of iterations of
the  OPE potential in the non-relativistic theory, this is no longer
possible when one is solving an integral equation - corresponding to an infinite number of iterations.   

In the present paper we consider the problem of three interacting
bosons at low energy. 
We identify the problem of non-uniqueness of the solution to the
integral equation at LO in the EFT as an artefact of the non-relativistic
approach and propose an alternative formulation based on 
 time-ordered perturbation theory (TOPT) applied to the
 Lorentz-invariant effective Lagrangian. 

\medskip

\begin{figure}[t]
\begin{center}
\epsfig{file=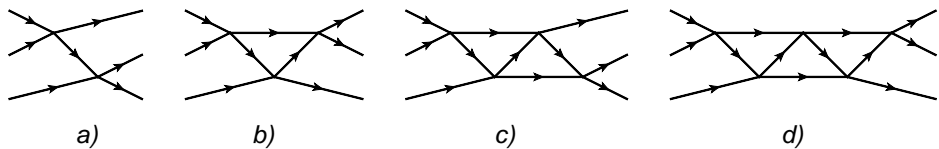,scale=0.75}
\caption{Examples of diagrams contributing to the particle-bound state scattering amplitude at leading order.}
\label{3BSc}
\end{center}
\end{figure}

 We consider a low-energy EFT of self-interacting scalar particles described  by an effective Lagrangian
\begin{equation}
\label{lagR} 
{\cal L}  = \frac{1}{2}\, \partial_\mu \phi(x) \partial^\mu \phi(x)
- \frac{1}{2}\,  m^2\,\phi^2(x) - \frac{\lambda}{4!} \,\phi^4(x) + {\cal L}_{ho}(x) 
\,,
\end{equation}
where $m$ and $\lambda $ are the mass of the scalar particle and the coupling constant of the LO interaction, respectively. 
The higher order Lagrangian ${\cal L}_{ho}$ contains an infinite number of  Lorentz-invariant self-interactions which are symmetric under $\phi\to -\phi$ transformations. 
To consider scattering of a scalar particle on a bound state of two
particles at low energies it is convenient to use the TOPT.  Standard
power counting considerations \cite{Weinberg:rz} suggest that at low
energies, for the case of large two-body scattering length,  the
leading contributions to the amplitude of three bosons going in three
bosons are given by diagrams, containing only particles (i.e. no
antiparticles) in intermediate states, examples of which are shown in
Fig.~\ref{3BSc}. For three-momenta of external particles of the
order of the inverse two-body scattering length, an
infinite number of such TOPT diagrams has to be summed up at LO.   
Note that the LO effective Lagrangian describes the standard,
perturbatively renormalizable $\phi^4$-theory and, therefore, none of
these diagrams require counter terms beyond the LO Lagrangian. 
On the other hand, different UV behaviour is observed in the
non-relativistic theory. In particular,  diagram b) is finite,
diagrams c) and d) are logarithmically and linearly divergent in
non-relativistic case. It is easily seen that adding one more loop
adds one power to the overall degree of divergence as it brings one
additional denominator, i.e.~two powers of momenta in the denominator,
and one additional three-dimensional integration. 
In general, more
iterations of the tree-order diagram a), considered as the kernel of an integral equation, 
bring higher- and higher-order overall divergences, 
which can be properly canceled only by including contributions of counter terms of higher and higher orders in momenta/energy. 
Thus, the non-relativistic problem, unlike its Lorentz-invariant counter part,  turns out to 
be perturbatively non-renormalizable. 
Notice that not every procedure that makes the physical quantities finite is a 
quantum field theoretical renormalization in EFT \cite{Epelbaum:2009sd}. 
In EFT, perturbative expansions of renormalized non-perturbative expressions must 
reproduce the renormalized perturbative series, wherever this expansion exists.
Therefore, a proper EFT renormalization of the three-body amplitude in a non-relativistic 
approach, which makes the non-perturbative expression cutoff independent and generates, 
if expanded, a perturbative series of subtracted diagrams, requires the inclusion of 
contributions of an infinite number of three-body counter terms.

\begin{figure}[tb]
\begin{center}
\epsfig{file=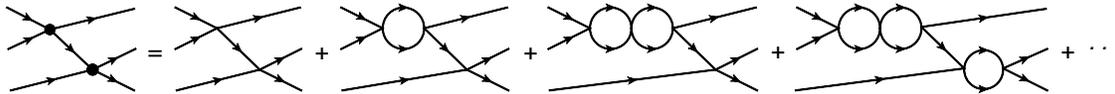,scale=0.90}
\caption{Examples of loop diagrams which sum up to a diagram with a two-body scattering amplitude as an effective vertex. effective vertex is indicated by a filled circle.}
\label{figRen}
\end{center}
\end{figure}

Subsets of the mentioned infinite number of diagrams, like ones shown
in Fig.~\ref{figRen}, sum up to diagrams where the four particle
interaction vertex is replaced by the scattering
amplitude.\footnote{Such resummed diagrams turn out to be finite also
  in the non-relativistic theory.}
For the purpose of this resummation, it is convenient to introduce a dummy field $T(x)$ with the quantum numbers
of two bosons \cite{kaplan97}, referred to as a ``dimeron''  \cite{Bedaque:1998km}, and consider 
\begin{eqnarray}
\label{lagtR} 
{\cal L}_T  &=&    \frac{1}{2}\, \partial_\mu \phi(x) \partial^\mu \phi(x)
- \frac{1}{2}\,  m^2 \, \phi^2(x) + \frac{\Delta}{2}\,T^2(x) - \frac{g}{\sqrt{2}} \,T(x)\, \phi^2(x)\,
\end{eqnarray} 
instead of the LO Lagrangian of Eq.~(\ref{lagR}). 
Here the scale parameter $\Delta$ is included to give the scalar field $T$ the usual
mass dimension.  The Lagrangian ${\cal L}_T$ describes exactly the same physics as the LO Lagrangian of Eq.~(\ref{lagR}).
The undressed dimeron propagator is a constant $i/\Delta$ and the
particle propagator is given by 
\begin{equation}
i\,S(k)  =  \frac{i}{k^2-m^2 +i\,0^+} = \frac{i}{2\omega(k) \left[ k_0-\omega(k)+i\,0^+\right]}+ \frac{-i}{2\omega(k) \left[ k_0+\omega(k)-i\,0^+\right]}
 \equiv i\,S_p(k) +i\,S_a(k),
\label{SRProp}
\end{equation}
where $\omega(k)=\sqrt{m^2+\vec k\,^2}$.  By using $S_p$ as the
propagator of the scalar particle in the pertinent Feynman diagrams and integrating
over the zeroth components of the loop momenta 
we end up with TOPT diagrams involving intermediate particle states only, i.e. we drop the TOPT diagrams with intermediate antiparticle states  which do not contribute at LO.

Denoting by $-i \,\Sigma$ the sum of one-particle irreducible diagrams contributing to the dimeron two point function, we obtain for the dressed dimeron propagator 
\begin{equation}
i S_\Delta(p) =  \frac {i}{\Delta - \Sigma}\,.
\label{ddpr}
\end{equation}
The leading one loop contribution to the dimeron self-energy has the form:
\begin{equation}
\label{Sigma}
\Sigma(p) =  g^2 I_{NN}(p^2)\,,
\end{equation}
where
\begin{equation}
I_{NN}\left( p^2 \right) = \frac{i}{(2 \pi)^4} \,\int
\,d^4k\,S_p(k)\,S_p(p-k)\,. \label{Jintegraldefinition}
\end{equation}
Performing renormalization by subtracting at threshold we obtain
\begin{equation}
\label{DpropDressed} 
i S_\Delta(p)\equiv i S_\Delta(p_0,\vec p) =  \frac {i}{\Delta^R -
g^{2}\,I^R_{NN}(p^2) +i\,0^+}\,,
\end{equation}
where 
$ I^R_{NN}\left( p^2 \right) =I_{NN}\left( p^2 \right)-I_{NN}\left( 4 m^2 \right)$ and
$\Delta^R$ is the renormalized parameter.

Attaching four boson lines to this dressed dimeron
propagator we get the two-particle scattering amplitude at LO:
\begin{equation}
-i\,A=\frac{-2 i g^2}{\Delta^R -
g^{2}\,I^R_{NN}(p^2) +i\,0^+}\,.
\label{2Bampl}
\end{equation}
Matching its non-relativistic expansion to the effective range expansion  we obtain $\Delta^R=\frac{g^2}{16 \pi a_2 m}$, where $a_2$ is the two-body scattering length. 

\begin{figure}[t]
\begin{center}
\epsfig{file=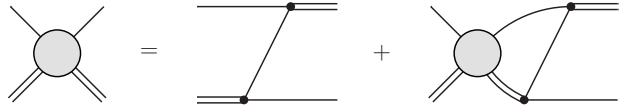,scale=0.55}
\caption{Graphical representation of the equation for the particle-dimeron scattering amplitude. Solid and doubled lines correspond to the particle and the dimeron, respectively.}
\label{figEQ}
\end{center}
\end{figure}

The amplitude of a scalar particle scattering off a bound state of two scalar particles can be obtained from the particle-dimeron scattering amplitude $T$. The LO contribution to the 
amplitude $T$ satisfies the equation graphically represented in Fig.~\ref{figEQ}.
Here, the incoming scalar particle and the dimeron are on-shell and have the momenta $(\omega(k),-\vec k \,)$ and $(\omega_B(k),\vec k\,)$, respectively. 
The outgoing particle and the dimeron are off-shell with momenta
$\left(\omega(k)-\epsilon,-\vec p\right)$ and
$\left(\omega_B(k)+\epsilon,\vec p\right)$, respectively, where  
$\omega_B(p)=\sqrt{(2m-B_2)^2+\vec p\,^2}$. Note that for better comparison with the 
non-relativistic work, we consider this half-off-shell amplitude. For a physical
scattering process, the final-state particles are, of course, on shell.
Thus, we have the following equation:
\begin{eqnarray}
i\, T(\vec k, \vec p,\epsilon) & = & -2 g^2 i\, S_p(\omega_B(k)-\omega(k)+\epsilon,\vec p+\vec k) \nonumber\\
&+& \int \frac{d^4 q }{(2 \pi)^4} i\, S_p(\omega(k)-\epsilon-q_0,-\vec q)\left[-2 g^2 i\, S_p(\omega_B(k)-\omega(k)+2 \epsilon+q_0,\vec p+\vec q)\right] \nonumber\\
&\times & i S_\Delta(\omega_B(k)+\epsilon+q_0,\vec q) \,i\,T(\vec k,\vec q,\epsilon+q_0).
\label{RSTM1}
\end{eqnarray}
Integration over $q_0$ leads to
\begin{eqnarray}
T(\vec k, \vec p) & = & \frac{- g^2}{\omega(k+p)\left[E-\omega(p)-\omega(k)-\omega(k+p)\right]}\nonumber\\
&+& 8 \pi \int \frac{d^3 \vec q }{(2 \pi)^3}\frac{T(\vec k, \vec q)}{-\frac{1}{a_2}+16 \pi m \,I^R_{NN}[P^2]}\,\frac{-m}{\omega(q)\omega(q+p) \left[E-\omega(p)-\omega(q)-\omega(q+p)\right]},
\label{RSTM}
\end{eqnarray}
where
\begin{equation}
E  =  \omega(k)+\omega_B(k),\ P^2 = \left[E-\omega(q)\right]^2-\vec q\,^2.
\label{defs}
\end{equation}

Next, to easily compare with the corresponding non-relativistic equation (7) of Ref.~\cite{Bedaque:1998km}, we define $t(\vec k, \vec p) = 2 m T(\vec k, \vec p) $ and obtain for the S-wave:
\begin{eqnarray}
t(k, p) & = & \frac{m g^2}{p k}\ln \frac{\omega(k)+\omega(p)+\sqrt{(k+p)^2+m^2}-E}{\omega(k)+\omega(p)+\sqrt{(k-p)^2+m^2}-E}\nonumber\\
&+&  \frac{2}{\pi} \int_0^\infty \frac{d q\, q^2 t(k, q) }{-\frac{1}{a_2}+J[P^2]}\,
\frac{m}{p q \,\omega(q)}\ln \frac{\omega(q)+\omega(p)+\sqrt{(q+p)^2+m^2}-E}{\omega(q)+\omega(p)+\sqrt{(q-p)^2+m^2}-E}\,,
\label{RSTMSPW}
\end{eqnarray}
where
\begin{eqnarray}
J[P^2] & = & 16 \pi m \, J^R_{NN}[P^2]= -\frac{m}{\pi P^2}\sqrt{P^2(P^2-4 m^2)} \ln\left[ 1-\frac{P^2}{2 m^2}+\frac{1}{2 m^2}\sqrt{P^2(P^2-4 m^2)}\right].
\label{integral}
\end{eqnarray}
Here, we have replaced the integrals $I^R_{NN}$ by $ J^R_{NN}\left( p^2 \right) =J_{NN}\left( p^2 \right)-J_{NN}\left( 4 m^2 \right)$, where 
\begin{equation}
J_{NN}\left( p^2 \right) = \frac{i}{(2 \pi)^4} \int
\,d^4k\,S(k)\,S(p-k) \label{integraldefinition}
\end{equation}
is the full Feynman integral  which differs from $I_{NN}$ by a higher-order contribution corresponding to a TOPT diagram with antiparticles and has an advantage that it can be calculated analytically. We emphasise that these two 
integrals have the same logarithmic ultraviolet behaviour for large momenta.

\begin{figure}[t]
\begin{center}
\epsfig{file=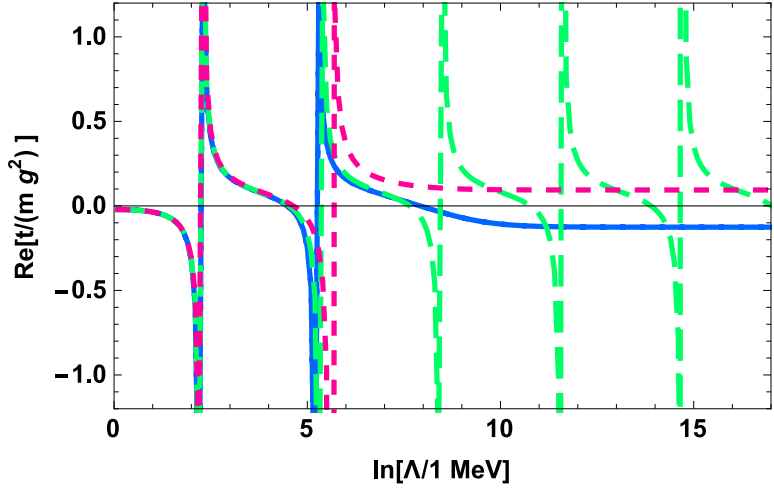,scale=0.80}
\caption{Cutoff dependence of the amplitude $t(k,p)/(m g^2)$  for $a_2=1\, {\rm MeV}^{-1}$, $m=939$~MeV, 
$k=0$~MeV and  $p=10$~MeV. The solid line corresponds to the solution
to the cutoff-regularized version of Eq.~(\ref{RSTMSPW}).  
The long-dashed and short-dashed lines correspond to the solution of cutoff-regularized Eq.~(\ref{STM}) and its modification by adding a factor of $1/(1+\ln(1+q/m))$ in the integrand, respectively. }
\label{figCD}
\end{center}
\end{figure}

Performing the standard $1/m$ expansion it is easily seen that Eq.~(\ref{RSTMSPW}) reduces to its non-relativistic analogue, Eq.~(7) of Ref.~\cite{Bedaque:1998km} 
with $\lambda=1$ (without three-boson contact interaction, i.e. for $h=0$) 
\begin{eqnarray}
t(k, p)  =  \frac{m g^2}{p k}\ln \frac{k^2+p^2+k p -m E_{\rm nr}}{k^2+p^2-k p -m E_{\rm nr}}
+  \frac{2}{\pi} \int_0^\infty \frac{d q\, q^2 t(k, q) }{-\frac{1}{a_2}+\sqrt{3 q^2/4-mE_{\rm nr}}}\,
\frac{1}{p q}\ln \frac{q^2+p^2+q p -m E_{\rm nr}}{q^2+p^2-q p -m E_{\rm nr}}\,,
\label{STM}
\end{eqnarray}
where $E_{\rm nr}=3 k^2/4 m-B_2$. Eq.~(\ref{STM}) is equivalent 
to the Skornyakov-Ter-Martirosyan  (S-TM) equation \cite{skornyakov}. 
Eq.~(\ref{STM}) does not have an unique solution \cite{danilov}.  By considering a cutoff-regularized equation one obtains a unique solution, however,  the removed cutoff limit does not exist \cite{Bedaque:1998kg}.
Equation  (\ref{RSTMSPW}) has a different (milder) ultraviolet behaviour. 
In particular, for large $q$ its integrand contains an additional factor of $\ln q$ in the denominator.  This factor plays a crucial role for the non-perturbative solution to Eq.~(\ref{RSTMSPW}). It is easily checked by numerical calculation  that 
the solution to the regularized equation, unlike to its non-relativistic analogue,  converges to a well defined limit when the cutoff, imposed on the integration, is taken to the infinity. The cutoff dependence of solutions to 
Eqs.~(\ref{RSTMSPW}) and (\ref{STM}) is shown in Fig.~\ref{figCD} where we plot $t(k,p)/(m g^2)$ as functions of $\Lambda$. 
While we give the results for the sharp cutoff, we have checked that smooth cutoff regulators lead to the same behaviour of the amplitudes.
We also plot the cutoff dependence of the modification of Eq.~(\ref{STM}) where we have included an additional factor of $1/(1+\ln(1+q/m))$ in the integrand. 
This simple modification clearly demonstrates that the logarithmic improvement of the UV behaviour leads to a cutoff-independent solution to the equation.

Relativistic corrections which are taken into account in the approach of this work induce an 
effective range of the two-body interaction $r_{\rm eff}\sim 1/m$ which is fixed by the 
mass of interacting particles. 
Its value is not related to the true range of the two-body interaction potential. 
This non-vanishing effective range guarantees that  the three-body
equation has a well-defined UV limit. However, in general, it cannot
describe three-body observables as they depend strongly on the range of the 
two-body potential for the case of a large two-body scattering
length.\footnote{This follows from the well-known Thomas collapse of the
  three-body system in the case of zero-range interactions \cite{Thomas:1935zz}.} In an EFT
with contact interactions only, 
all exchange particles are integrated out and the range of the interaction is encoded 
in the contact interaction 
terms involving derivatives acting on fields. The non-perturbative inclusion of such two-body interaction terms with derivatives in the calculations of the two-body scattering amplitude 
can be handled by applying the BPHZ subtractive renormalization similarly to Refs.~\cite{Gegelia:1998gn,Gegelia:1998iu}. However, we are unable to extend this scheme to 
the three-body sector. A natural solution to the problem is provided by the inclusion of the exchange particles as dynamical degrees of freedom. In case of EFT 
for nucleons, the inclusion of the pions in Lorentz-invariant
formulations with TOPT leads to perturbatively renormalizable and
non-perturbatively finite results both in the $^1S_0$ and $^3S_1-^3D_1$ partial waves. 
For nucleon-deuteron scattering in the doublet channel the resulting system of coupled
integral equations is perturbatively renormalizable, and we expect
that non-perturbatively it has a finite unique solution. Work along
this line is in progress. 
Thus, we argue that meaningful
predictions in the 3-body system with a large two-body scattering
length for the case with $a_2 \gg r_2 \gtrsim m^{-1}$ 
require non-perturbative inclusion of the range ``correction''
(i.e. the range of the two-body interaction $r_2$),  while for $r_2\ll m^{-1}$ the perturbative 
treatment of the range corrections should be adequate. On the other hand, while the three-body interaction
is not required by the UV renormalization, we do not see a possibility
to make an a-priori estimation of its actual impact on low-energy
observables.
It might be negligible for one physical system while its effect for another system might be so large that the perturbative treatment of the three-body interaction does not lead to a satisfactory description of the data - 
such a case would pose a challenge to our renormalizable approach. This is because the three-body interaction is perturbatively non-renormalizable
already at LO, therefore its non-perturbative inclusion does not seem to be
feasible in the framework with the removed-cutoff limit. 
A possible solution to this problem is provided by invoking a finite ultraviolet cutoff
\cite{Epelbaum:2008ga}.

\medskip

\acknowledgments

The authors thank H.-W. Hammer and X.-L. Ren for the comments on the manuscript.
This work was supported in part by Georgian Shota Rustaveli National
Science Foundation (grant FR/417/6-100/14), DFG (SFB/TR 110,
``Symmetries and the Emergence of Structure in QCD''), BMBF (contract No. 05P2015 -
NUSTAR R\&D) and the National Science Foundation (Grant No.~NSF PHY11-25915). 
The work of UGM was supported in part by The Chinese Academy of Sciences 
(CAS) President's International Fellowship Initiative (PIFI) grant no. 2015VMA076.


\begin{references}
 
\bibitem{Bedaque:1998kg} 
  P.~F.~Bedaque, H.~W.~Hammer and U.~van Kolck,
  Phys.\ Rev.\ Lett.\  {\bf 82}, 463 (1999).


\bibitem{Bedaque:1998km} 
  P.~F.~Bedaque, H.~W.~Hammer and U.~van Kolck,
  Nucl.\ Phys.\ A {\bf 646}, 444 (1999).

\bibitem{Bedaque:1999ve} 
  P.~F.~Bedaque, H.~W.~Hammer and U.~van Kolck,
  Nucl.\ Phys.\ A {\bf 676}, 357 (2000).

\bibitem{Blankleider:2000vi} 
  B.~Blankleider and J.~Gegelia,
  nucl-th/0009007.

\bibitem{Braaten:2004rn} 
  E.~Braaten and H.-W.~Hammer,
  Phys.\ Rept.\  {\bf 428}, 259 (2006).

\bibitem{Bernard:1992qa} 
  V.~Bernard, N.~Kaiser, J.~Kambor and U.-G.~Mei{\ss}ner,
  Nucl.\ Phys.\ B {\bf 388}, 315 (1992).

\bibitem{Gegelia:1999gf} 
  J.~Gegelia and G.~Japaridze,
  Phys.\ Rev.\ D {\bf 60}, 114038 (1999).

\bibitem{Epelbaum:2012ua}
  E.~Epelbaum and J.~Gegelia,
  Phys.\ Lett.\ B {\bf 716}, 338 (2012).

\bibitem{Weinberg:rz}
S.~Weinberg,
Phys.\ Lett.\ B {\bf 251}, 288 (1990).

\bibitem{Epelbaum:2009sd} 
  E.~Epelbaum and J.~Gegelia,
  Eur.\ Phys.\ J.\ A {\bf 41}, 341 (2009).

\bibitem{kaplan97}
 D.B. Kaplan, {\it Nucl. Phys.} {\bf B494} (1997) 471;

\bibitem{skornyakov}
    G.~V.~Skornyakov and Ter-Martirosyan,
  Sov.\ Phys.\ JETP {\bf 4}, 648 (1957)   [Zh.\ Eksp.\ Teor.\ Fiz.\  {\bf 31}, 775 (1956)].

\bibitem{danilov}
 G.S. Danilov, { Sov. Phys. JETP} {\bf 13}, 349 (1961).

\bibitem{Thomas:1935zz} 
  L.~H.~Thomas,
  Phys.\ Rev.\  {\bf 47}, 903 (1935).

\bibitem{Gegelia:1998gn} 
  J.~Gegelia,
  Phys.\ Lett.\ B {\bf 429}, 227 (1998).

\bibitem{Gegelia:1998iu} 
  J.~Gegelia,
  J.\ Phys.\ G {\bf 25}, 1681 (1999).

\bibitem{Epelbaum:2008ga} 
  E.~Epelbaum, H.~W.~Hammer and U.-G.~Mei\ss ner,
  Rev.\ Mod.\ Phys.\  {\bf 81}, 1773 (2009).



\end{references}
\end{document}